# Dodecacene Generated on Surface:

# Re-opening of the Energy Gap

*Frank Eisenhut,[1,2,+] Tim Kühne,[1,2,+] Fátima García,[3,+] Saleta Fernández,[3] Enrique Guitián,[3] Dolores Pérez,[3] Georges Trinquier,[6] Gianaurelio Cuniberti,[2,4] Christian Joachim,[5] Diego Peña,[*,3] and Francesca Moresco[*,1]*

[1]Center for Advancing Electronics Dresden, TU Dresden, 01069 Dresden, Germany

[2]Institute for Materials Science, TU Dresden, 01069 Dresden, Germany

[3]Centro de Investigación en Química Biolóxica e Materiais Moleculares (CiQUS) and Departamento de Química Orgánica, Universidade de Santiago de Compostela, Santiago de Compostela 15782, Spain

[4]Dresden Center for Computational Materials Science (DCMS), TU Dresden, 01069 Dresden, Germany

[5]GNS & MANA Satellite, CEMES, CNRS, 29 rue J. Marvig, 31055 Toulouse Cedex, France

[6]Laboratoire de Chimie et Physique Quantiques, IRSAMC-CNRS-UMR5626, Université Paul-Sabatier (Toulouse III), 31062 Toulouse Cedex 4, France.

*Correspondence should be addressed to diego.pena@usc.es and francesca.moresco@tu-dresden.de.

ABSTRACT

The acene series represents a unique model system to investigate the intriguing electronic properties of extended π-electron structures in the one-dimensional limit, which are important for applications in electronics and spintronics and for the fundamental understanding of electronic transport. Here we present the on-surface generation of the longest acene obtained so far: Dodecacene. Scanning tunneling spectroscopy gives access to the energy position and spatial distribution of its electronic states on the Au(111) surface. We observe that, after a progressive closing of the gap and a stabilization to about 1 eV at the length of decacene and undecacene, the energy gap of dodecacene unexpectedly increases to 1.4 eV. Considering the acene series as an exemplary general case, we discuss the evolution with length of the single tunneling resonances in comparison with ionization energy, electronic affinity, and optical gap.

Acenes are polycyclic aromatic hydrocarbons formed by linearly fused benzene rings. They have been intensively studied in the past mainly because of their application for low band gap and high electronic mobility materials[1]. The recent renewal interest in acenes is motivated by their expected higher radical character as well as the stabilization of the optical excitation energy for an increasing number of fused rings. These two features make extended acenes and their derivatives very attractive for nano-electronic devices requiring a low (but non-zero) electronic band gap and magnetic properties in the ground state.

Acenes with more than five rings are chemically very reactive, unstable to ambient conditions, and have low solubility. Spectroscopy of their low-energy excitations has been therefore mainly restricted to optical absorption spectra obtained from matrix isolation studies[2, 3, 4]. On-surface generation is a new powerful method to obtain acenes *via* the rational synthesis of precursor molecules[5], making possible to study higher acenes by high-resolution scanning probe techniques. The on-surface investigation of *single* acenes is therefore developing as a promising complementary investigation method for exploring the high reactivity of higher acenes and unravelling their properties[5, 6, 7, 8, 9, 10, 11, 12, 13, 14].

In recent works, we have presented the variation of the energy gap of higher acenes up to decacene as a function of the number N of linearly fused benzene rings, combining experimental results with theoretical considerations[8] and showing that the energy gap between the first electronic tunneling resonances below and above the Fermi energy remains finite up to this length. We have theoretically explained this effect by a di-radical perturbative contribution to the electronic ground state, completing the closed shell singlet ground state $|S0>$[8].

New works published in 2019 are continuing this fundamental discussion, indicating that the debate on the electronic structure of higher acenes is actual and far from being solved [12, 15, 16]. In particular, the group of Roman Fasel discusses a possible open-shell character of the ground state already for N = 9 (nonacene) on Au(111)[12]. The open-shell character is motivated not only by the evolution of the gap, but also by the high reactivity of N = 9 (nonacene) and is experimentally observed by the formation of complexes with single Au adatoms. Furthermore, in a monoelectronic approximation, the groups of F. Evers and R. Korytár[15] predict incommensurate size-oscillations of the energy gap of oligoacenes adsorbed on Au(111). The authors suggest that the experimental observation of such oscillations should be possible for oligoacenes with slightly more than 10 rings adsorbed on a gold surface. On the other hand, S. Hemmatiyan and D.A. Mazziotti[16] in a theoretical study observe that the calculation of the band gap for the acene series requires the accurate treatment of the electronic correlations, and predict for acenes of increasing length a finite gap without oscillations. By comparing these three exemplary publications and our work[8], it is evident that further experiments are needed to draw clear conclusions on the electronic structure of oligoacenes. The next step is therefore the on-surface synthesis of dodecacene reported in this article.

Here, we present the on-surface generation of N = 12 (dodecacene), the larger acene generated to date, and the study of its electronic properties by scanning tunneling microscopy and spectroscopy. We observe that, after a progressive closing of the gap and its stabilization to about 1 eV for N = 10 (decacene) and N = 11 (undecacene), the energy gap at N = 12 (dodecacene) surprisingly increases again to 1.4 eV. Considering the acene series as an exemplary general case for our understanding of the electronic structure of a homo-electronic series of molecules, we discuss the evolution of the single tunneling resonances with the number of fused rings in comparison with ionization energy, electronic affinity, and optical gap.

**RESULTS**

**On-surface generation of dodecacene**

First, we designed and synthesized stable dodecacene precursors by solution chemistry. Based on our previous experience on the generation of decacene[7], we decided to prepare pentaepoxy derivative **1**, (Fig 1a) a molecule with six Clar sextets in its structure. Dodecacene precursor **1** was obtained by a four-steps iterative sequence of aryne cycloadditions in solution (see supporting information for details on the synthesis). Then, to obtain highly reactive dodecacene (**3**), with only one Clar sextet in its closed-shell resonance structure, we deposited the pentaepoxy derivative **1** on a Au(111) surface kept at room temperature .

We obtained a submonolayer coverage of the partially deoxygenated dodecacene precursors **2** with two epoxy groups still attached. We induced an on-surface deoxygenation by annealing the sample to 220°C. After the annealing, dodecacene molecules and a few partially deoxygenated precursors were visible on the surface, as shown in the overview STM image of Fig. 1b.

A partially deoxygenated dodecacene precursor **2** with two remaining epoxy groups is visible in the STM image of Fig. 2a, and clearly identified by the superposition with the chemical structure of the precursor in Fig. 2b. It is important to note that, as already reported for the on-surface formation of heptacene from the corresponding precursor[8], the epoxy groups are visible in a STM image only at high bias voltage (V =-1 V in this case). The last step of the precursor deoxygenation leading to the formation of dodecacene on surface can be induced by inelastic tunneling electrons, as shown in Fig. 2c. Fig. 2d presents a high-resolved STM image of a single dodecacene molecule recorded at constant height mode with a CO-functionalized tip. The image clearly shows the twelve lobes corresponding to the twelve benzene rings of dodecacene, proving the formation of dodecacene on Au(111) by the on-surface reduction of the epoxy precursors. In summary, the reaction can be either induced thermally by annealing the surface at T > 220°C, or partial thermally and concluded by inelastic tunneling electrons.

**Electronic structure of dodecacene**

To further investigate the electronic structure of N = 12 (dodecacene), we performed scanning tunneling spectroscopy (STS) experiments on the individual molecules. A typical differential conductance spectrum (dI/dV) is presented in Fig. 3. We observe five tunnel electronic resonance peaks that we call for simplicity: R-2, R-1, R0, R1, and R2. Similar spectra have been reported for smaller acenes $3 \leq N \leq 11$[8, 10] (and are compared in the Supplementary Information).

In Fig. 4, differential conductance maps recorded at the corresponding resonances are shown. The maps visualize the spatial distribution of the molecular resonances and are very similar to those obtained for N = 10 (decacene)[8], showing however the features of the two additional rings. The number of rings unambiguously verifies the assignment of the reacted molecule to N = 12 (dodecacene). Similar to the case of the shorter acenes, the variations of spatial differential conductance on those maps are mainly determined by single electron transfer contributions to the tunneling current, from the tip to the surface through the molecule (or *vice versa*). They are therefore well reproduced by calculating the conductance map of the molecular orbitals in a mono-electronic approximation[8].

The energy of the resonance peaks maxima in Fig. 3 are in good agreement with the energies at which the differential conductance maps show well-resolved resonance profiles (see Supplementary Information). We can therefore assign R-2 to the energy of -1120 mV, R-1 to -620 mV, R0 to -320 mV, R1 to 1070 mV, and R2 to 1850 mV. An uncertainly of about 100 meV should be considered taking into account the energy involved in the surface screening of the metal surface for the physisorbed molecule, and reflected on the finite width of the peaks of Fig. 3.

To study the evolution of the observed electronic resonances and energy gap with the acene length, we have plotted the energy position of the single resonances (Fig. 5a) and of the (R1 – R0) gap (Fig. 5b) as a function of N. As one can see, after a saturation at about 1 eV at the length of N = 11 (undecacene), the gap re-opens to about 1.4 eV for N = 12 (dodecacene). By considering separately the evolution of the resonances R0 and R1 shown in Fig. 5a, one can observe that while R0 remains nearly constant for N ≥ 8, R1 (as well as R2) clearly increases from undecacene to dodecacene. As a practical consequence, we are not able to observe the third resonance R3 of dodecacene in the voltage range available for STS measurements, which is roughly between -3 V and +3 V.

**DISCUSSION**

For the interpretation of our experimental results, we have first identified the observed tunneling resonances and considered their relation to ionization potentials and optical transitions of the known oligoacenes. To this aim, we have plotted in Fig. 6 the experimental optical frontier electronic resonances relative to the vacuum level of Au(111) (and therefore renormalized considering the Au(111) work function of 5.31 eV). We have also plotted the variations with N of the first ionization energy (IE1), the electronic affinity (EA), and the optical gap ($E_g$).

**The R0 resonance and the ground state**

Optically, the energy position of the ground state can be assigned using the first ionization energy (IE1). IE1 is experimentally known for oligoacenes with N ≤ 7[17], as shown by the black curve in Fig. 6 (IE1 exp). Although the theoretical IE1 curve is available for N ≤ 11[17], we have recalculated it for N ≤ 13 at B3LYP/6-311G** level,[18] obtaining the green curve in Fig. 6 (IE1 calc). The small systematic shift between the experimental (black curve) and the DFT-calculated

IE1 (green curve) has been already noticed,[19, 20, 21] and, in this specific case, it might originate in the increasing contribution of di-radical states reaching a quantum weight of 20 % to 30 % of the total ground state for N = 15[22].

The blue curve in Fig. 6 shows the variation of the resonance R0 with N recorded by scanning tunneling spectroscopy. This curve nicely follows the calculated IE1 (green curve), also for the case of N = 12 (dodecacene). Therefore, we assign the R0 resonance to the polyacene ground state. It is important to note that this ground state is neither the closed shell singlet |S0> nor a di-radical state. It is rather a quantum superposition between the neutral |S0> and the radical forms[8]. The conductance map recorded at R0 is the signature of the mono-electronic contribution of the highest occupied molecular orbital (HOMO) to this ground state, since both the |S0> and the di-radical states can be decomposed on Slaters determinants with a dominant weight for the acene HOMOs.

**The R1 resonance and the first excited state**

The next step is to identify the STS resonance R1 of N = 12 (dodecacene). For this purpose, we consider first the optical gap ($E_g$) that has been recently measured for acenes with N ≤ 11[4]. For comparison, we plot in Fig.6 (pink curve) ($E_g$ + R0) as a function of N. This is slightly artificial because of the mixture of STS measurements (electronic) with optical measurements.

However, we can consider it valid because we have just assigned R0 to the ground state resonance. By comparing now the energy position of this first optical excited state with the STS resonance R1 (brown curve in Fig. 6), we can clearly see that both curves follow the same trend. Therefore, we can assign in a first approximation the R1 resonance to the energy position of the |S1> excited state of polyacene. This is in agreement with the fact that, similar to the case of R0, the conductance map recorded at R1 is the signature of the mono-electronic contribution of the lowest unoccupied molecular orbital (LUMO) to the first singlet excited state, since |S1> can be decomposed on Slaters determinants with a dominant weight for the acene LUMOs.

Fig. 6 also evidently shows that the R1 resonance does not correspond to the electronic affinity (EA). EA (known experimentally up to N = 7)[17] is plotted in red in Fig. 6 for comparison. The confusion comes from the fact that R1 results from an elementary electron transfer process, which is described at this energy by a Slater determinant with a virtual occupation of the LUMO without depopulating the HOMO. This virtual state is sometimes mistakenly interpreted as a reduced state of the molecule. However, quantum mechanically, the occupation of this state at the resonance cannot be more than 50 % and happens only during a short electron transfer event (few fs). As a result, R1 it is not exactly |S1>. We will further call it in this way for convenience. However, this fact can thereafter influence the interpretation of the R1 resonance of dodecacene discussed below.

**The case of dodecacene**

Fig. 5 noticeably shows a re-opening of the energy gap for N = 12, mainly due to the shift of R1 to higher energies and not significantly determined by R0. This important result provides further relevant information to the actual scientific debate on the intriguing electronic structure of oligoacenes.

Starting from Ref. [23] and [15], the increasing of the gap for N = 12 can be treated as an example of a gap oscillation due to the presence of a 'Dirac cone' in the band structure, leading to the oscillation of the R0 and R1 resonances with N. However, such oscillations exist only in the frame of a Bloch-like band structure theory based on a mono-electronic approximation[8]. Contrary to this interpretation and as recently discussed in the literature[8, 16], the calculation of the band gap of oligoacenes requires a careful treatment of the electron correlations and leads to the absence of gap oscillations with no gap closure while increasing N. In particular, for N = 10, a contamination of 38 % by doubly excited configurations prevents the R0−R1 gap to reach zero *via* the stabilization of R0 as calculated up to N = 12$^8$. An oscillation of the gap, and the related validity of the mono-electronic approximation, can be therefore excluded in the case of oligoacenes.

A different interpretation of the re-opening of the gap for N = 12 could be a charging of N = 12 (dodecacene) on Au(111). As one can see in Fig. 6, R0 is approaching the Fermi level of Au(111) by increasing N, allowing eventually a partial charge transfer from the molecule to this surface. This charge transfer would depopulate the ground state with an impact on the weight of its di-radical contamination, and a consequent destabilization of R1 and R2 via a shifting of the excited states. Notice that in this case the ionization cannot be complete, otherwise the observed STS tunneling spectrum would be strongly modified. For example, the ground state R0 of dodecacene(+) would be much lower in energy than observed here (ref. $^{17}$).

Following this interpretation, it would remain however unclear why the increase of the R1 energy takes place for dodecacene and not already for shorter acenes, being the ionization energy already very close to the Fermi level at N = 8.

We prefer to explain the re-opening of the gap for N = 12 by a destabilization of the electronic resonance R1 (and R2) through higher molecular empty states. This effect is probably related to the tunneling process via virtually occupied empty states and does not necessarily correspond to the optical transition process to the same empty states. As discussed above, while increasing N, the tunneling process via |S1> cannot be described anymore by a simple virtual occupation of the LUMO. For acenes with large N, the R1 resonance involves a combination of electron transfer paths through the virtual (and instantaneous) quantum occupations of LUMO+1, LUMO+2, etc. This can lead to a shift of the R1 (and R2) resonance with a consequent re-opening of the R0-R1 gap, while R0 remains stable.

The poly-radical character of acenes has been shown to increase with their length N, with the first di-radical appearing around N=7 and the tetra-radical around N=12[24]. This interpretation, which can be understood in a simplified way from Clar's aromatic sextet rule[25, 26], is confirmed in Fig. 6 by a relative maximum of the R1 (and $E_g$ + R0) curve at N = 7, followed by a minimum around N = 10, and a maximum at N = 12.

**CONCLUSIONS**

In conclusion, the on-surface synthesis of unprecedented dodecacene presented in this article represents a fundamental step to understand the electronic properties of a one-dimensional conjugated π-system, polyacene, as a function of its length. High-resolution STM images and STS differential conductance maps unambiguously demonstrate the successful deoxygenation of the epoxy precursors and the formation of dodecacene on the Au(111) surface.

While we can assign the R0 resonance to the polyacene ground state, the unexpected increase of the energy of the empty-state tunneling electronic resonances and the consequent reopening of the gap suggest us to reconsider the state of the art explanations of the acenes electronic structure, pointing out to a possible increase of the poly-radical character of the empty-state electronic resonances at the Clar sextets N=7 and N=12.

**METHODS**

**Experimental Details.** STM experiments were performed using a custom-built instrument operating at a low temperature of T = 5 K and ultra-high vacuum (p ≈ 1 × 10$^{-10}$ mbar) conditions. All shown STM images were recorded in constant-current mode with the bias voltage applied to the sample. Differential conductance spectra were measured using lock-in detection with a modulation frequency of 833 Hz and a modulation amplitude of 20 mV. Only those tips showing the Au(111) surface state[27] at -0.51 V were used for spectroscopy measurements on the molecules.

The precursor molecules (**1**, see supporting information for details on the synthesis) were deposited at a submonolayer coverage on the clean Au(111) surface kept at room temperature by flash heating from a Si wafer, as described in Ref. [28]. This method prevents the decomposition of the precursors during sublimation.

CO molecules were deposited with a very low surface coverage onto the cold sample and then deliberately picked up by the tip of the STM to functionalize the apex.[29]

**Calculations:** The IE1 green triangles from N = 4 to N = 13 in Fig. 6 were calculated at B3LYP/6-311G** level using the Gaussian-09 package[18]. These points correspond to vertical ionization energies from closed-shell ground states.

## ASSOCIATED CONTENT

**Supporting Information.** Synthetic details, NMR data of the precursors, and additional STS data are available free of charge via the Internet at http://pubs.acs.org.

## AUTHOR INFORMATION

**Corresponding Authors**

*Correspondence should be addressed to diego.pena@usc.es and francesca.moresco@tu-dresden.de.

**Author contributions**

All authors discussed the results and contributed to the manuscript. The manuscript was written through contributions of all authors. All authors have given approval to the final version of the manuscript. [+]These authors contributed equally.

**ACKNOWLEDGEMENTS**

Discussion with J.P. Malrieu from the Laboratoire de Chimie et Physique Quantiques, IRSAMC-CNRS-UMR5626 are deeply acknowledged. This work was funded by the European Union's Horizon 2020 research and innovation program under the project MEMO (grant agreement No. 766864) and the project SPRING (grant Agreement No. 863098). Support by the German Excellence Initiative via the Cluster of Excellence EXC1056 ''Center for Advancing Electronics Dresden'' (cfaed) is acknowledged. We thank financial support by the Agencia Estatal de Investigación (MAT2016-78293-C6-3-R and CTQ2016-78157-R), Xunta de Galicia (Centro singular de investigación de Galicia, accreditation 2016-2019, ED431G/09), and the Fondo Europeo de Desarrollo Regional (FEDER). F.G. acknowledges the Juan de la Cierva Incorporación 2017 program.

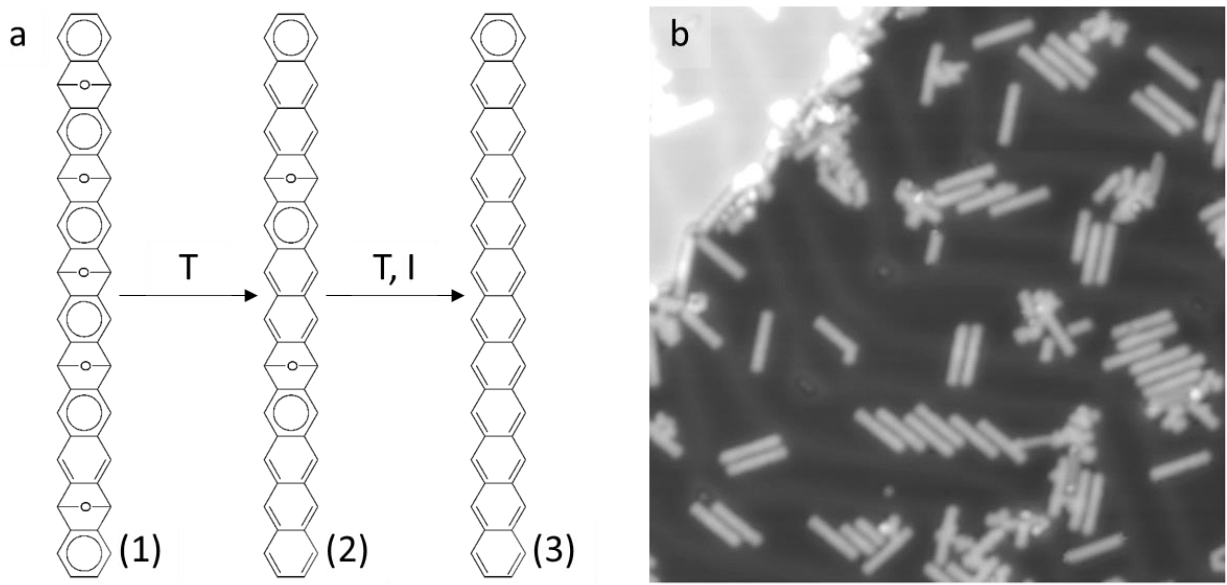

**Figure 1.** Dodecacene precursors and on-surface reaction. (a) Chemical structure of the precursor and schematic representation of the on-surface reaction (1) Structure of the dodecacene precursor, characterized by five epoxy groups; (2) Partially deoxygenated precursor showing two remaining epoxy groups; (3) final dodecacene molecule. (b) STM topographical overview showing dodecacene and partially deoxygenated precursor molecules on Au(111) after annealing the sample to 220°C. Image parameters: Vbias = 500 mV, I = 9 pA, size 33 x 33 nm2.

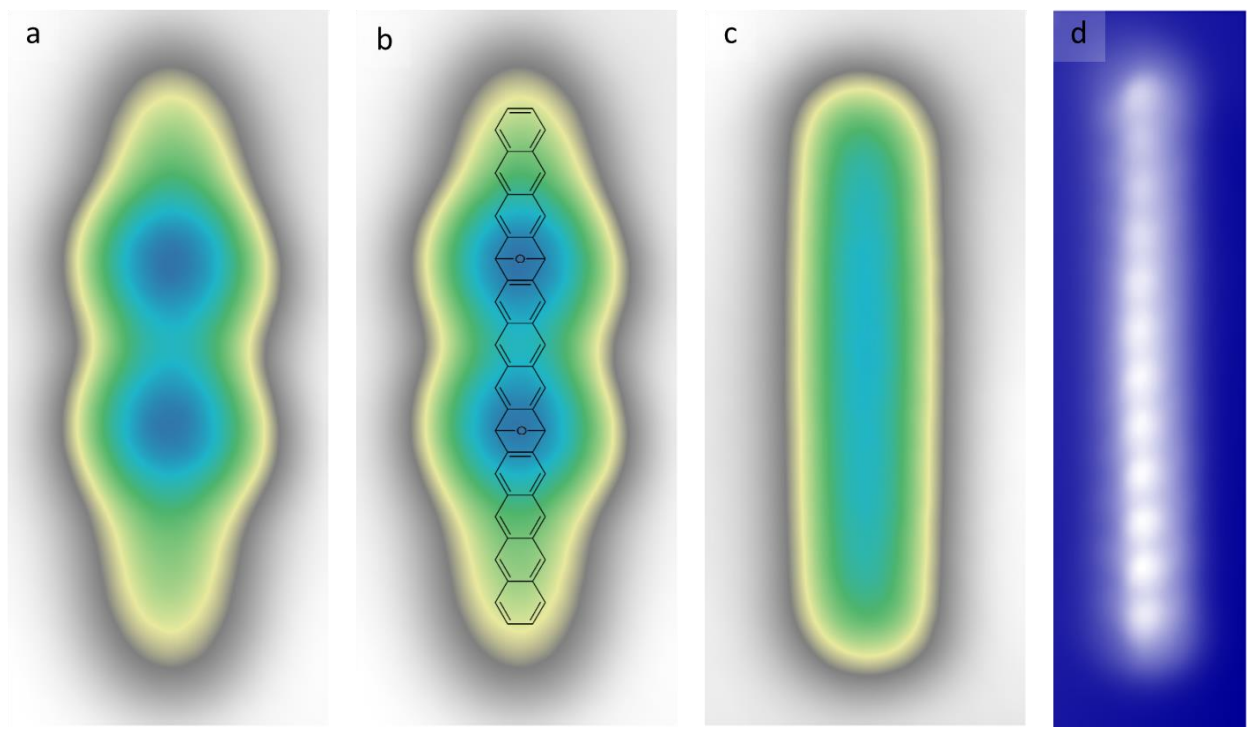

**Figure 2.** STM images of dodecacene on Au(111). (a) STM image of dodecacene with two remaining epoxy groups. Image parameters: $V_{bias}$ = 2 V I = 13 pA, image size 2 nm x 4.5 nm. (b) Superposition of the STM image of (a) with the structure of the partially deoxygenated precursor identifying the position of the two epoxy groups. (c) STM image of a fully deoxygenized dodecacene. Image parameters: $V_{bias}$ = -1 V, I = 10 pA, image size 2 nm x 4.5 nm. (d) Constant height STM image using a CO-functionalized tip, recorded at $V_{bias}$ = 10 mV. 12 bright lobes, representing the twelve benzene rings, are well distinguishable. Image size: 1 nm x 4 nm.

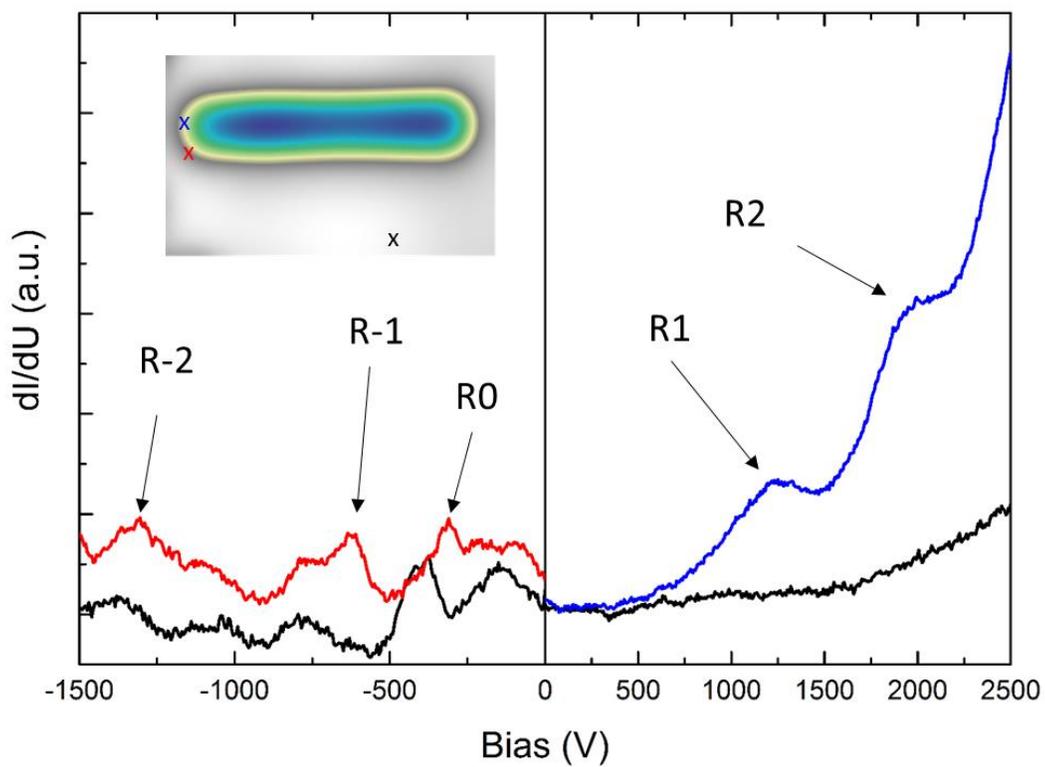

**Figure 3.** STS dI/dV spectrum of dodecacene. Spectra were recorded at the positions shown in the STM image of the inset. Five resonances are observed on the molecule, that we call for simplicity: R-2, R-1, R0, R1, and R2.

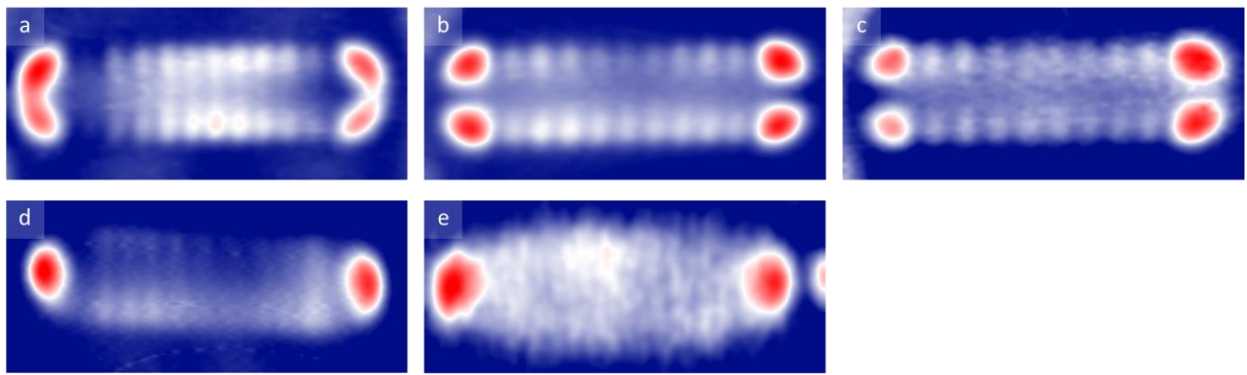

**Figure 4.** dI/dV maps of the dodecacene electronic resonances. Image size: 4.7 nm x 2 nm. (a) R0 at V = -320 mV; (b) R-1 at V = -620 mV; (c) R-2 at V = -1220 mV; (d) R1 at V = 1070 mV; (e) R2 at V = 1815 mV.

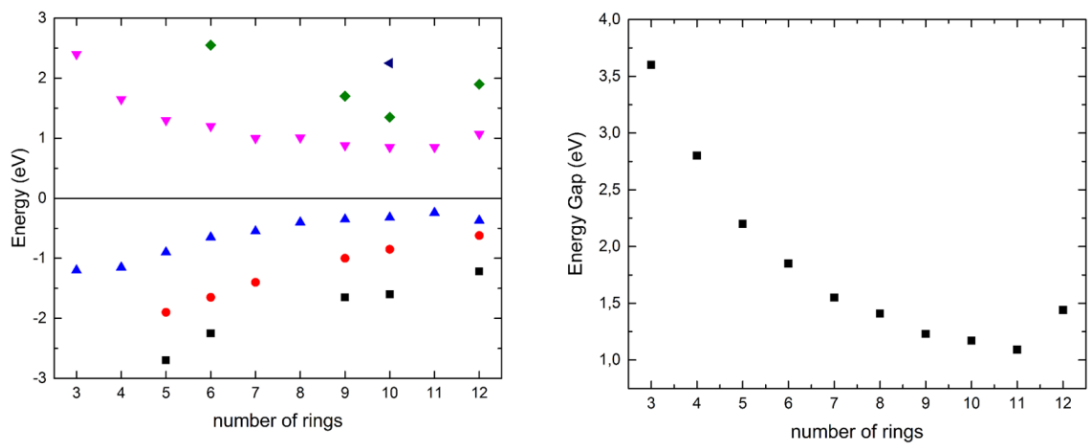

**Figure 5.** Evolution of resonance energies and gap. (a) Evolution of the energy of the single resonances from anthracene to dodecacene. Black squares: R-2, red circles: R-1, blue triangles: R0, pink triangles: R1, green diamonds: R2, dark blue triangle R3. The values for anthracene to undecacene taken from ref.[6]. (b) Development of the energy gap of acenes dodecacene as a function of the number of benzene rings. The energy gap is converging to a value of around 1 eV until undecacene. For dodecacene, the energy gap increases again to about 1.4 V.

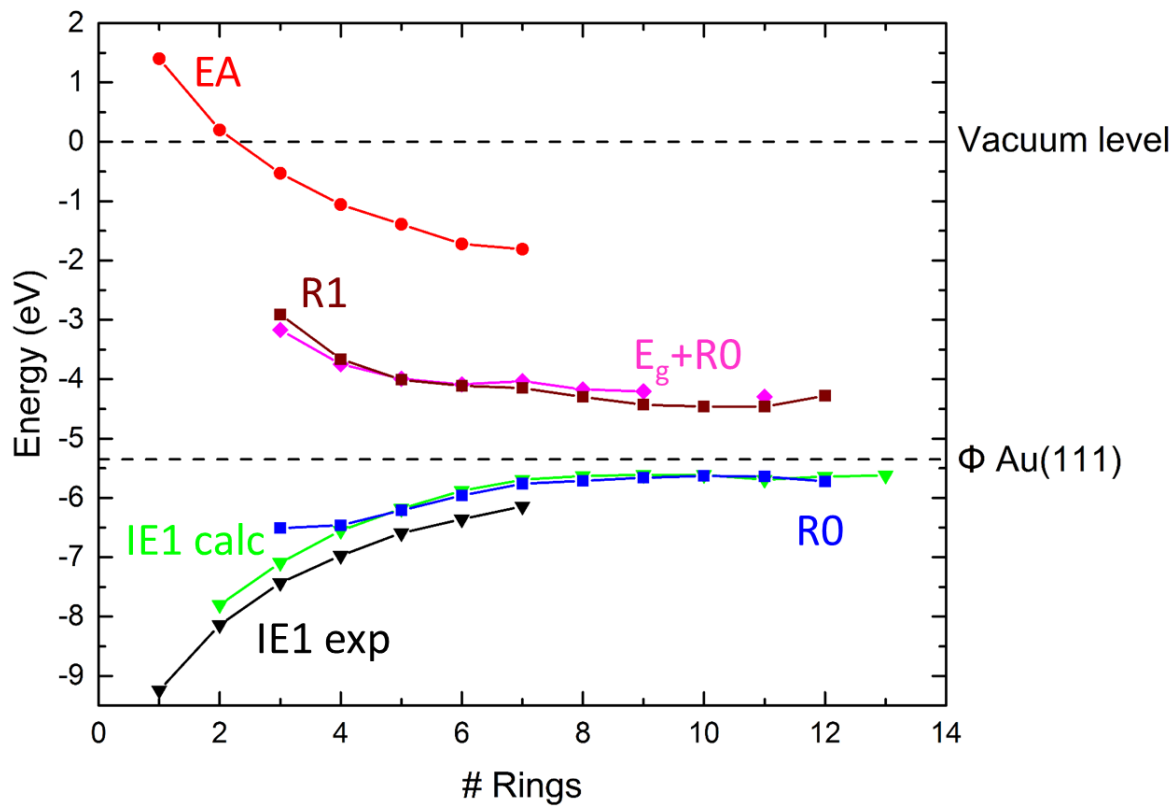

**Figure 6.** Evolution with length of the single tunneling resonances in comparison with ionization energy, electronic affinity, and optical gap. Black triangles: experimental ionization energy from ref. [17] (IE1 exp). Green triangles: calculated ionization energy (IE1 calc). Blue squares: resonance R0 recorded by STS relative to the Au(111) vacuum level (renormalized considering the Au work function of 5.31 eV) (R0). Braun squares: resonance R1 recorded by STS relative to the Au(111) vacuum level (R1). Pink diamonds: optical transition considering the optical $E_g$[4]; The plotted values are given by summing the optical gap to R0 ($E_g$ + R0). Red circles: electronic affinity (EA)[17].

## Dodecacene Generated on Surface: Re-opening of the Energy Gap

*Frank Eisenhut,*[1,2,+] *Tim Kühne,*[1,2,+] *Fátima García,*[3,+] *Saleta Fernández,*[3] *Enrique Guitián,*[3] *Dolores Pérez,*[3] *Georges Trinquier,*[6] *Gianaurelio Cuniberti,*[2,4] *Christian Joachim,*[5] *Diego Peña,**[,3] *and Francesca Moresco**[,1]

[1]Center for Advancing Electronics Dresden, TU Dresden, 01069 Dresden, Germany

[2]Institute for Materials Science, TU Dresden, 01069 Dresden, Germany

[3]Centro de Investigación en Química Biolóxica e Materiais Moleculares (CiQUS) and Departamento de Química Orgánica, Universidade de Santiago de Compostela, Santiago de Compostela 15782, Spain

[4]Dresden Center for Computational Materials Science (DCMS), TU Dresden, 01069 Dresden, Germany

[5]GNS & MANA Satellite, CEMES, CNRS, 29 rue J. Marvig, 31055 Toulouse Cedex, France

[6]Laboratoire de Chimie et Physique Quantiques, IRSAMC-CNRS-UMR5626, Université Paul-Sabatier (Toulouse III), 31062 Toulouse Cedex 4, France.

*Corresponding authors. E-mail: diego.pena@usc.es; francesca.moresco@tu-dresden.de

[+]These authors contributed equally to this work.

# Table of Contents

## 1      Synthetic details

**General methods for the synthesis of the precursors**

All reactions were carried out under argon using oven-dried glassware. $CH_2Cl_2$ and $CH_3CN$ were dried using a MBraun SPS-800 Solvent Purification System. Finely powdered CsF was purchased from ABCR GmbH and dried under vacuum at 100 °C, cooled under argon and stored in a glove-box. Furan was purchased from Sigma-Aldrich and filtered through neutral aluminium oxide (Brockmann I) and stored with molecular sieves (3 Å) under argon atmosphere. Other commercial reagents were purchased from ABCR GmbH, Sigma-Aldrich or Acros Organics, and were used without further purification. Deuterated solvents were purchased from Acros Organics. TLC was performed on Merck silica gel 60 $F_{254}$ and chromatograms were visualized with UV light (254 and 365 nm) and/or stained with Hanessian's stain. Column chromatography was performed on Merck silica gel 60 (ASTM 230-400 mesh). $^1$H and $^{13}$C NMR spectra were recorded at 300 and 75 MHz (Varian Mercury-300 instrument) or 500 and 125 MHz (Varian Inova 500 or Bruker 500) respectively. APCI high resolution mass spectra were obtained on a Bruker Microtof. The synthesis of the dodecacene precursor (**1**) was made by the following route (Scheme S1).

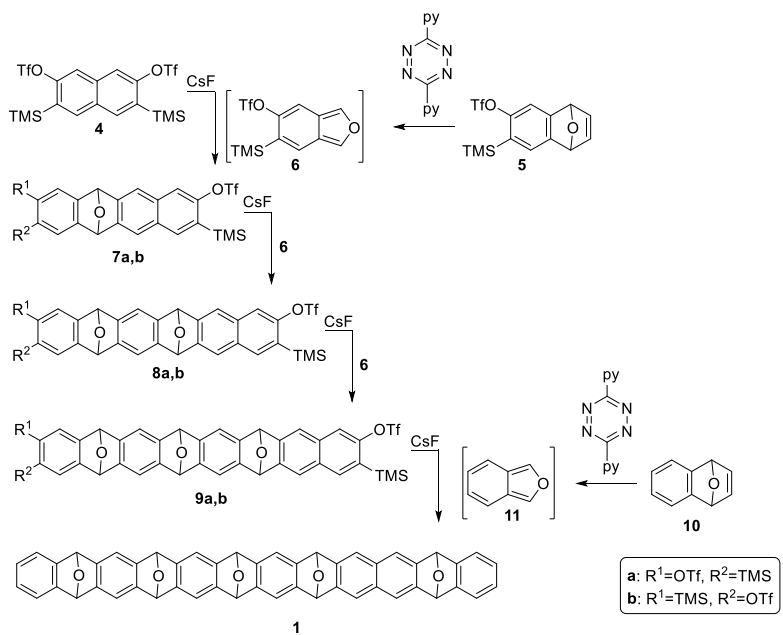

**Scheme S1.** Synthesis of dodecacene precursor **1**.

Compounds **4**,[i] **5**,[ii] **7a,b**,[2] **8a,b**[iii] and **10**[iv] were prepared following reported procedures and showed the same spectroscopic properties as reported therein.

**Synthesis of bistriflate 9a,b**

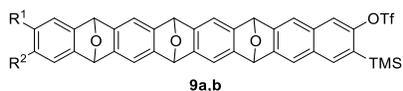

Compound **5** (102 mg, 0.28 mmol), and 3,6-di(pyridin-2-yl)-1,2,4,5-tetrazine (66 mg, 0.28 mmol) were dissolved in 5 mL of dry $CH_2Cl_2$ under Ar and stirred at 45 ºC. The reaction was monitored by TLC until complete consumption of **5** was observed (1.5 hours). Then, the reaction mixture was allowed to warm up to room temperature and the aryne precursor **8a,b** (150 mg, 0.088 mmol) dissolved in MeCN (3 mL) and CsF (19 mg, 0.124 mmol) were added. The reaction mixture was stirred at 45 ºC for 16 hours. Then, the solvent was evaporated under reduced pressure and the crude purified by chromatographic column (silica gel, hexane/$CH_2Cl_2$/diethyl ether 1/1/0.5) isolating 14 mg of **9a,b** as a mixture of regio and diastereomers. Yellow solid. Yield: 8%.

$^1$H NMR (500 MHz, CDCl$_3$) δ 7.83-7.77 (singlet group, 1H), 7.63-7.58 (singlet group, 3H), 7.33-7.30 (s, 2H), 7.27 (s, 2H), 7.22 (s group, 2H), 6.11- 6.08 (singlet group, 2H), 5.91 - 5.88 (singlet group, 4H), 0.35, 0.35, 0.18, 0.17. ppm.

$^{19}$F NMR (282 MHz, CDCl$_3$) δ -74.25, -74.46, -74.99, -75.09.

$^{13}$C NMR (126 MHz, CDCl$_3$) δ 153.1 (C Ar), 152.6 (C Ar), 152.6 (C Ar), 152.3 (C Ar), 152.2 (C Ar), 147.7 (C Ar), 147.6 (C Ar), 147.6 (C Ar), 147.6 (C Ar), 147.5 (C Ar), 147.5 (C Ar), 147.4 (C Ar), 147.3

(C Ar), 146.8 (C Ar), 146.7 (C Ar), 146.4 (C Ar), 146.4 (C Ar), 146.0 (C Ar), 145.9 (C Ar), 145.8 (C Ar), 145.8 (C Ar), 145.8 (C Ar), 145.8 (C Ar), 145.8 (C Ar), 145.1 (C Ar), 145.1 (C Ar), 137.1 (C Ar), 137.0 (C Ar), 135.7 (C Ar), 133.4 (C Ar), 133.3 (C Ar), 133.2 (C Ar), 131.1 (C Ar), 131.1 (C Ar), 130.8 (C Ar), 130.8 (C Ar), 130.5 (C Ar), 129.6 (C Ar), 129.6 (C Ar), 129.4 (C Ar), 129.4 (C Ar), 128.2 (C Ar), 126.5 (C Ar), 126.4 (C Ar), 125.5 (C Ar), 119.7 (C Ar), 119.7 (C Ar), 119.5 (C Ar), 119.5 (C Ar), 118.5 (CH Ar), 118.5 (CH Ar), 118.4 (CH Ar), 118.4 (CH Ar), 117.2 (C Ar), 117.2 (C Ar), 117.0 (CH Ar), 117.0 (C Ar), 116.8 (C Ar), 114.3 (CH Ar), 114.2 (CH Ar), 114.1 (CH Ar), 114.0 (CH Ar), 113.9 (CH Ar), 113.8 (CH Ar), 113.8 (CH Ar), 113.8 (CH Ar), 112.7 (CH Ar), 112.5 (CH Ar), 82.4 (CH-O), 82.4 (CH-O), 82.2 (CH-O), 82.2 (CH-O), 82.1 (CH-O), 82.0 (CH-O), 81.9 (CH-O), -0.7 ($CH_3$-TMS), -0.8, -0.9 ($CH_3$-TMS), -1.0 ($CH_3$-TMS), -1.0 ($CH_3$-TMS).

HRMS (APCI): for $C_{42}H_{35}F_6O_9S_2Si_2$ [M+1] calculated 917.1160, observed 917.1159.

**Synthesis of the dodecacene precursor**

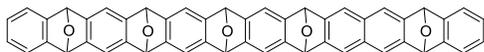

4-Dihydro-1,4-epoxynaphthalene (**10**) (43 mg, 0.30 mmol), and 3,6-di(pyridin-2-yl)-1,2,4,5-tetrazine (70 mg, 0.30 mmol) were dissolved in 2.5 mL of dry $CH_2Cl_2$ under Ar atmosphere and stirred at 45 ºC. The reaction was monitored by TLC until complete consumption of **10** was observed (30 min). Then, highly reactive isobenzofuran **11** was isolated by a short and fast column chromatography (hexane/$CH_2Cl_2$ 1/1). Isobenzofuran **11** was added immediately in anhydrous $CH_2Cl_2$ solution (3 mL) after isolation to a solution of **9a,b** in anhydrous MeCN (3 mL). Afterwards, CsF was added and the reaction mixture was stirred at 40ºC for 16 h. Then, the solvent was evaporated under reduced pressure and the crude was purified by chromatographic column (silica gel, hexane/ethyl acetate/diethyl ether 1/1/0.5) isolating two fractions of **1**: the less polar fracction (1 mg) and the more polar fraction (3 mg) as white solids. Combined yield 37%.

*Less polar fraction*: A mixture of two diastereomers in a 74:26 ratio. NMR data is given for the mayor isomer.

$^1$H NMR (500 MHz, CDCl$_3$) δ 7.50 – 7.47 (m, 2H), 7.34 (s, 2H), 7.33 – 7.30 (m, 2H), 7.25 (s, 2H), 7.23 – 7.20 (m, 2H), 7.15 (s, 2H), 7.03 – 7.00 (m, 2H), 6.97 – 6.93 (m, 2H), 6.07 (s, 2H), 5.99 (s, 2H), 5.86 (s, 2H), 5.79 (d, *J* = 1.5 Hz, 2H), 5.77 (s, 2H).

$^{13}$C NMR (126 MHz, CDCl$_3$) δ 148.0 (C Ar), 147.5 (C Ar), 147.3 (C Ar), 147.3 (C Ar), 147.1 (C Ar), 146.9 (C Ar), 146.3 (C Ar), 144.9 (C Ar), 144.3 (C Ar), 130.9 (C Ar), 126.2 (CH Ar), 125.7 (CH Ar), 120.3 (CH Ar), 120.1 (CH Ar), 119.1 (CH Ar), 119.1 (CH Ar), 113.7 (CH Ar), 82.5 (CH-O), 82.4 (CH-O), 82.2 (CH-O), 82.2 (CH-O), 82.1 (CH-O).

HRMS (APCI): for $C_{50}H_{29}O_5$ [M+1] calculated 709.2010, observed 709.2009.

*More polar fraction*: A mixture of at least 3 different diastereomers. It was not possible to assign the signals; therefore, NMR data is given without integration. This fraction was used for on-surface experiments.

1H NMR (500 MHz, Chloroform-d) δ 7.58 (s), 7.48 (s), 7.48 (s), 7.42 (s), 7.38 (m), 7.37 – 7.35 (m), 7.33 (s), 7.28 – 7.27 (m), 7.26 (s), 7.21 (s), 7.19 (s), 7.18 (s), 7.17 (s), 7.07 (s), 7.06 – 7.04 (m), 6.99 (m), 6.92 (m), 6.87 (m), 6.72 (m), 6.69 (m), 6.18 (s), 6.14 (s), 6.08 (s), 6.02 (s), 5.95 (s), 5.95 (s), 5.83 (s), 5.82 (s), 5.81 (s), 5.80 (s), 5.79 (s), 5.77 (s), 5.76 (s), 5.72 (s), 5.68 (s), 5.35 (m).

13C NMR (126 MHz, CDCl$_3$) δ 148.4 (C Ar), 147.2 (C Ar), 146.9 (C Ar), 146.7 (C Ar), 146.3 (C Ar), 146.1 (C Ar), 144.7 (C Ar), 144.1 (C Ar), 130.8 (C Ar), 130.8 (C Ar), 126.7 (CH Ar), 126.6 (CH Ar), 126.3 (CH Ar), 126.2 (CH Ar), 125.8 (CH Ar), 125.6 (CH Ar), 120.3 (CH Ar), 120.1 (CH Ar), 120.0 (CH Ar), 119.5 (CH Ar), 119.4 (CH Ar), 119.2 (CH Ar), 119.0 (CH Ar), 119.0 (CH Ar), 118.9 (CH Ar), 114.0 (CH Ar), 113.9 (CH Ar), 113.7 (CH Ar), 113.6 (CH Ar), 113.5 (CH Ar), 82.5 (CH-O), 82.4 (CH-O), 82.3 (CH-O), 82.2 (CH-O), 82.1 (CH-O), 81.9 (CH-O), 77.2 (CH-O).

HRMS (APCI): for C$_{50}$H$_{29}$O$_5$ [M+1] calculated 709.2010, observed 709.2011.

## 2   NMR data

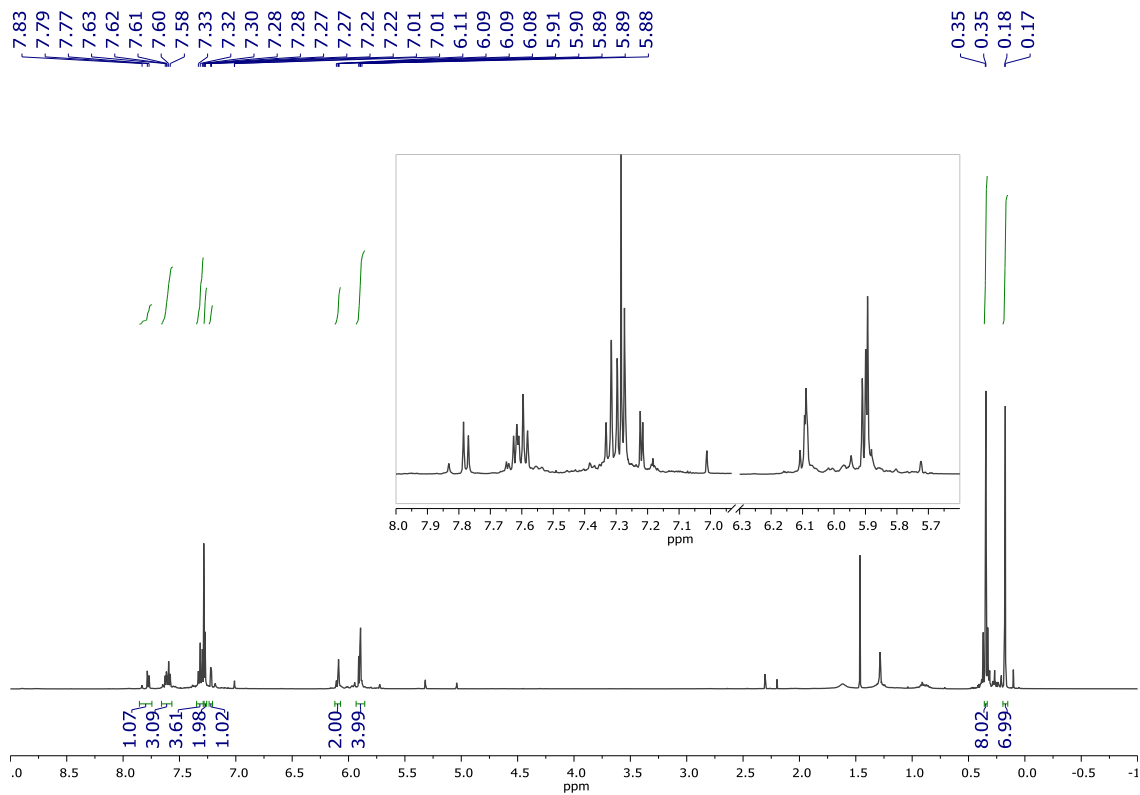

**Figure S1.** $^1$H NMR (500 MHz, CDCl$_3$) of **9a,b**.

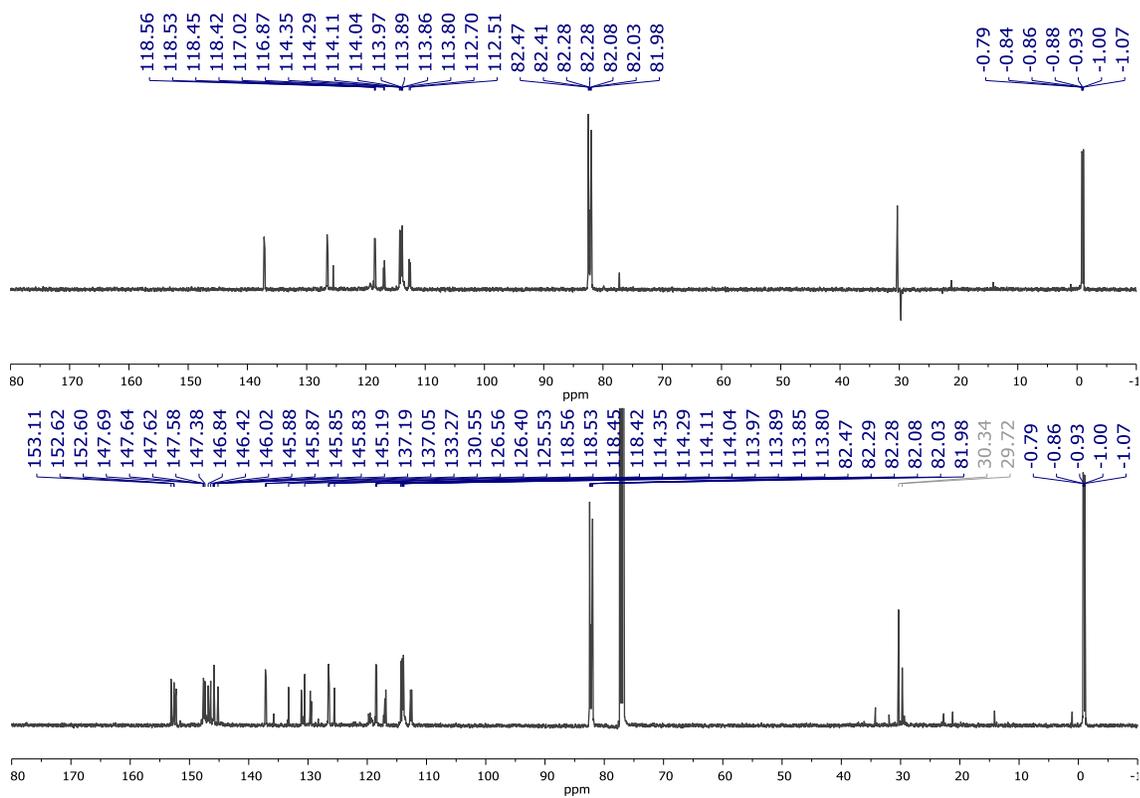

**Figure S2.** $^{13}$C (bottom) NMR (126 MHz, CDCl$_3$) and DEPT 135 (top) of **9a,b**.

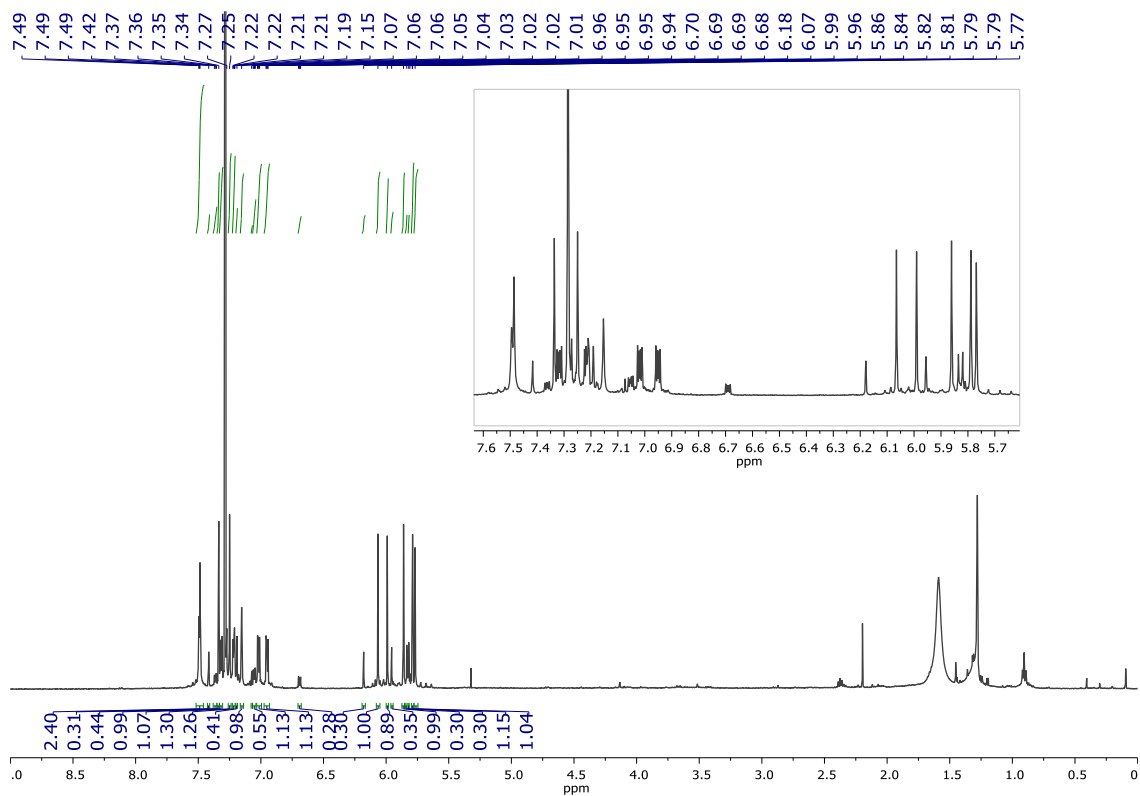

**Figure S3.** $^1$H NMR (500 MHz, CDCl$_3$) of the less polar fraction of dodecacene precursor **1**.

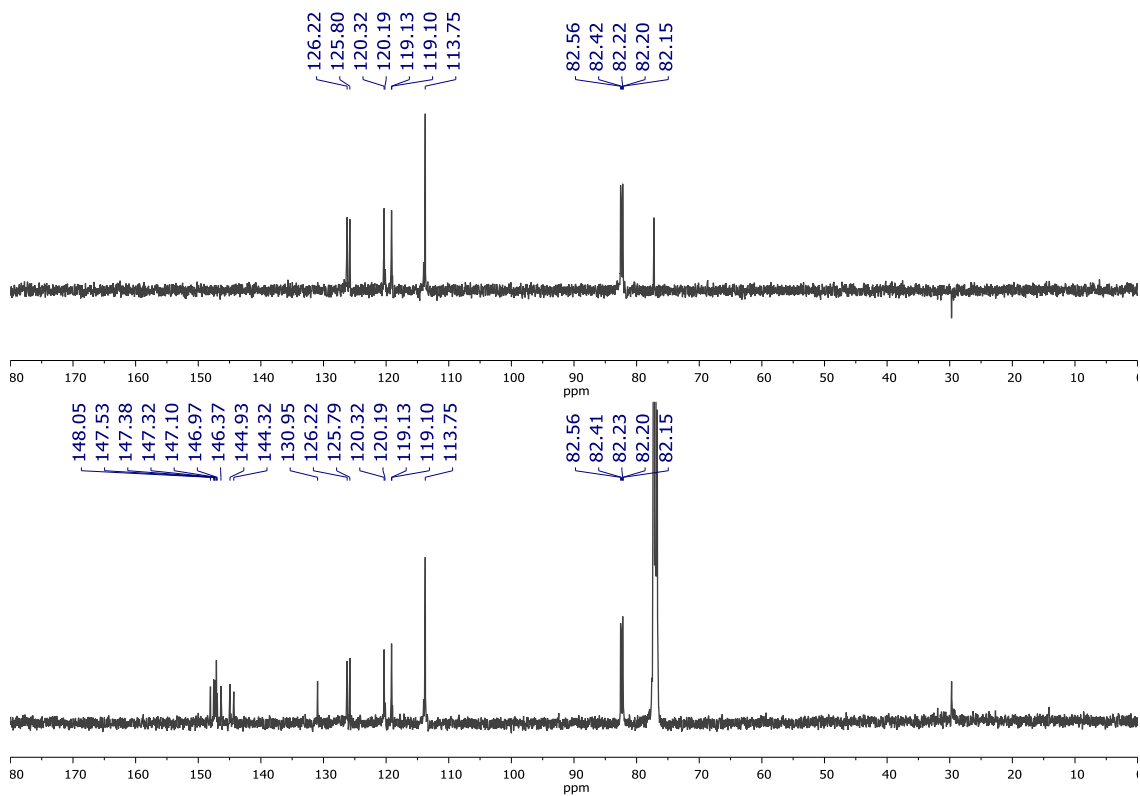

**Figure S4.** [13]C (bottom) NMR (126 MHz, CDCl₃) and DEPT 135 (top) of the less polar fraction of dodecacene precursor **1**.

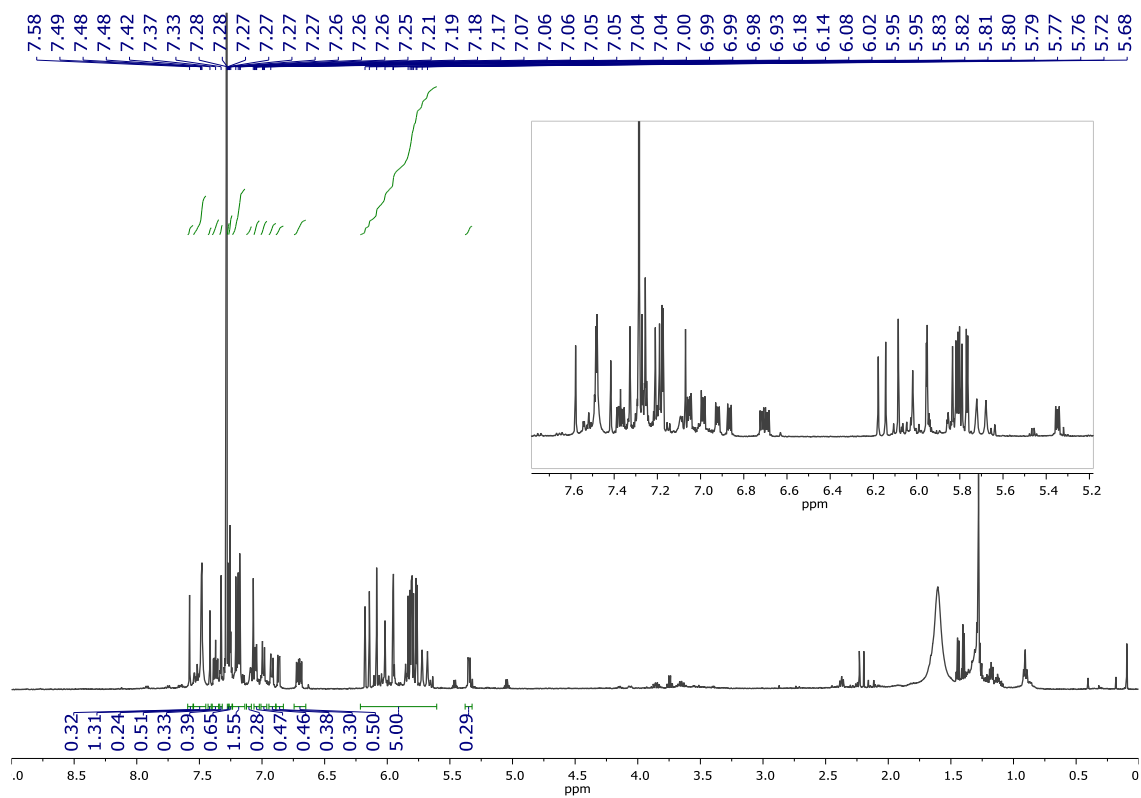

**Figure S5.** [1]H NMR (500 MHz, CDCl₃) of the more polar fraction of dodecacene precursor **1**.

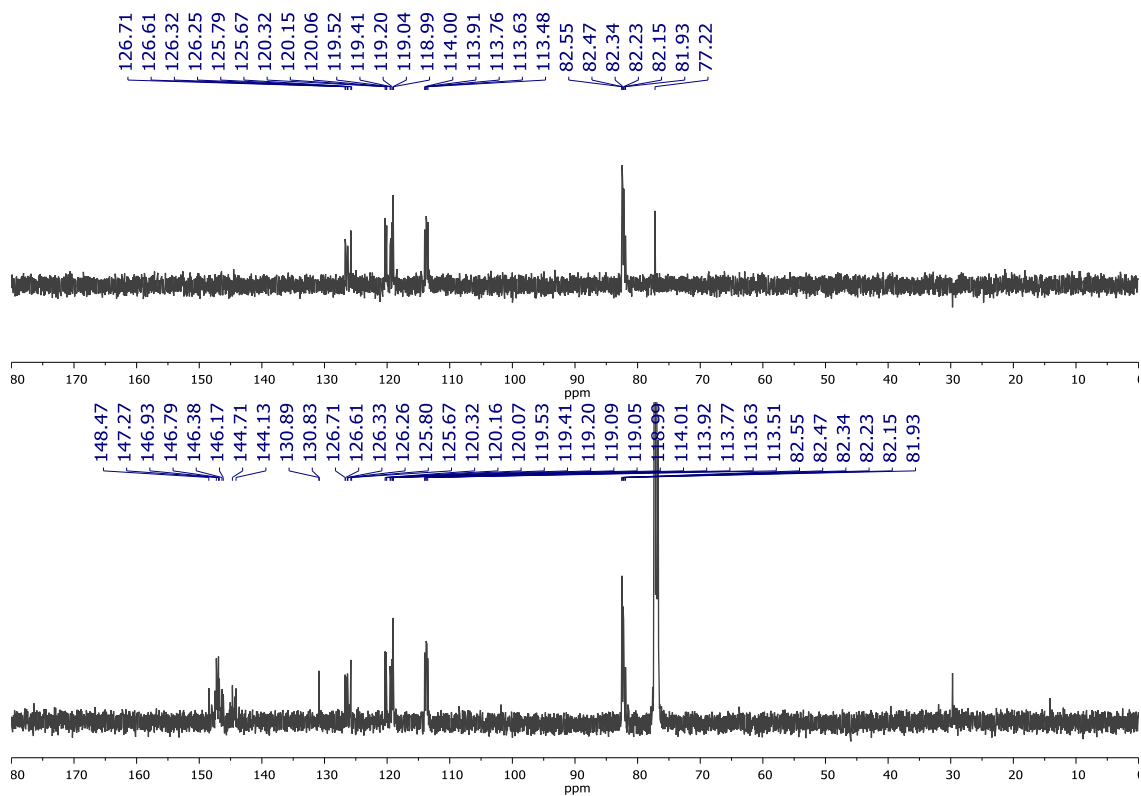

**Figure S6.** $^{13}$C (bottom) NMR (126 MHz, CDCl$_3$) and DEPT 135 (top) of the more polar fraction of dodecacene precursor **1**.

## 3   Additional STS data

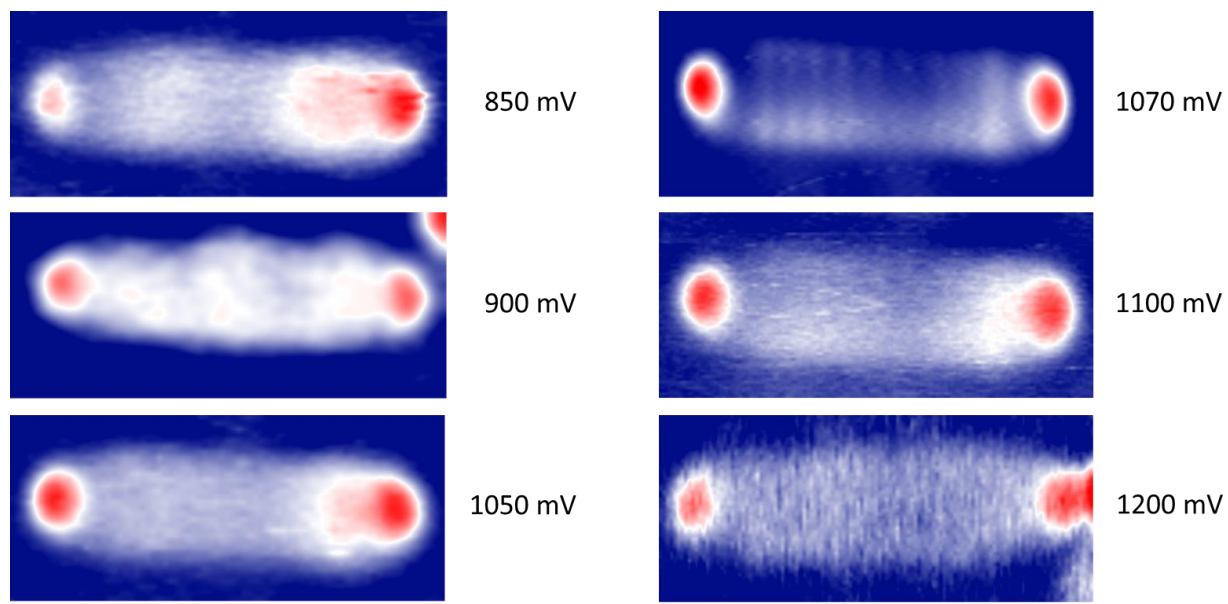

**Figure S7.** dI/dV maps of the R1 resonance of dodecacene measured at different bias voltage. The evolution of the contrast of the resonance features can be clearly followed and shows a maximum of contrast for V = 1070 mV with the typical LUMO contrast.

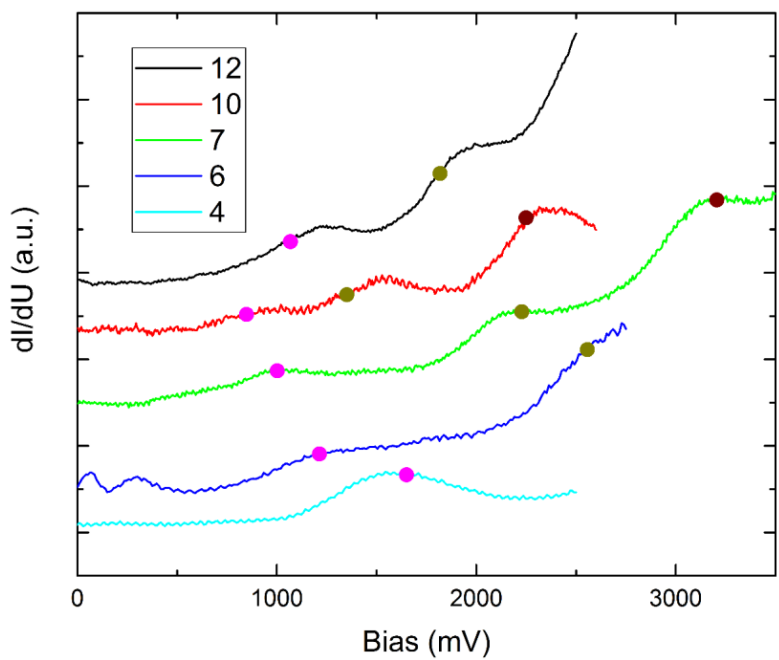

**Figure S8.** Spectra of the tunneling resonances of the unoccupied states for different numbers of fused benzene rings in the acene molecules, measured by scanning tunneling spectroscopy (STS). For each plot, the energy corresponding to the dI/dV map with the best contrast is marked: R1(pink), R2(gold), R3(brown). As one can see, the best resolved dI/dV maps (corresponding to the highest contribution of the monoelectronic component) do not exactly correspond to the maximum of the resonance peaks.